\documentclass[pra,amsmath,twocolumn,superscriptaddress]{revtex4}

\usepackage{graphicx}
\usepackage{amssymb}

\begin{document}

\title{Mechanisms of two-color laser-induced field-free molecular orientation}

\author{Michael Spanner}
\affiliation{Steacie Institute for Molecular Sciences, National Research Council of Canada, Ottawa, ON, Canada K1A 0R6}

\author{Serguei Patchkovskii}
\affiliation{Steacie Institute for Molecular Sciences, National Research Council of Canada, Ottawa, ON, Canada K1A 0R6}

\author{Eugene Frumker}
\affiliation{Joint Attosecond Science Laboratory, University of Ottawa and National Research Council of Canada, 100 Sussex Drive, Ottawa, On, Canada}
\affiliation{Department of Physics, Texas A\&M University, College Station, Texas 77843, USA}
\affiliation{Max-Planck-Institut f\"ur Quantenoptik, Hans-Kopfermann-Strasse 1, D-85748 Garching, Germany}

\author{Paul Corkum}
\affiliation{Joint Attosecond Science Laboratory, University of Ottawa and National Research Council of Canada, 100 Sussex Drive, Ottawa, On, Canada}

\begin{abstract}
Two mechanisms of two-color ($\omega$ + $2\omega$) laser-induced field-free
molecular orientation, based on the hyperpolarizability and ionization depletion,
are explored and compared.  The CO molecule is used as a
computational example.  While the hyperpolarizability
mechanism generates small amounts of orientation at intensities below the
ionization threshold, ionization depletion quickly becomes the dominant
mechanism as soon as ionizing intensities are reached. Only the ionization
mechanism leads to substantial orientation (e.g. on the order of
$\langle\cos\theta\rangle \gtrsim 0.1$).  For intensities typical of
laser-induced molecular alignment and orientation experiments, the two mechanism
lead to robust, characteristic timings of the field-free orientation
wave-packet revivals relative to the the alignment revivals and the revival
time.  The revival timings can be used to detect the active orientation mechanism
experimentally.
\end{abstract}

\date{\today}

\maketitle

\begin{figure}[b]
	\centering
	\includegraphics[width=0.8\columnwidth]{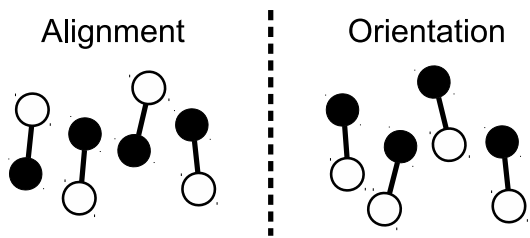}
	\caption{Cartoon illustrating the difference between alignment and orientation
	of a schematic polar diatomic molecule.}
	\label{FigAlignOrient}
\end{figure}

Laser-induced field-free alignment of small gas phase molecules, where the
molecular axis is aligned along particular direction (see
Fig.\ref{FigAlignOrient}), is now routine \cite{StapelfeldtSeideman} and is a
quickly becoming a central tool to in attosecond \cite{AttoReview} and
photoionization \cite{StolowAlign} experiments.  Much less studied is the
laser-induced field-free orientation of polar molecules, where both the
molecular axis and the asymmetry point along a particular direction, which has
recently been achieved \cite{Kling} using two-color ultrafast pulses built
from a fundamental and its second harmonic.  The underlying physical effect
thought to be responsible for the orientation is the hyperpolarizability
interaction \cite{Kling,Hyper}.  However, it is possible that a second
mechanism -- ionization depletion -- is operative, where a two-color ultrafast
pulse selectively ionizes molecules at particular angles with respect to the
polarization direction of the laser field.  Hyperpolarizability generates
orientation by causing an asymmetrical force that pushes the molecules toward
orientation, while ionization depletion generates orientation by directly
carving out an asymmetrical angular distribution upon ionization.  In this work
we implement a simple model of the ionization depletion mechanism, and contrast
and compare the two mechanisms.  A rigid linear rotor is used as an example
system. Extension to symmetric and asymmetric tops is conceptually
straightforward.  Rotor parameters corresponding to the CO molecule are used in
numerical examples.


The two-color laser pulse is written as
\begin{equation}\label{EqEt}
	E(t) = E_0 f(t) [\cos(\omega t) + \cos(2\omega t) ],
\end{equation}
where $f(t)$ is the pulse envelope ($0 \leq f(t) \leq 1$), and $E_0$ (a
positive real number) controls the peak electric field strength.  The relative
phase of the two fields was set to zero, and their relative amplitudes were set
to one, in order to maximize the field asymmetry that leads to orientation.
The nuclear rotation effectively does not respond on the timescale of the
carrier oscillations, and it is then appropriate to use the cycle-averaged
rotational Hamiltonian of the system:
\begin{equation}\label{EqHamil}
	H(\theta,t) = BJ(J+1) + V_P(\theta,t) + V_H(\theta,t) + V_I(\theta,t),
\end{equation}
where $B$ is the rotational constant, $V_P(\theta,t)$ is the polarizability
term that generates molecular alignment \cite{Herschbach}, $V_H(\theta,t)$ is
the hyperpolarizability term, and $V_I(\theta,t)$ accounts for ionization (see
below). All equations use Hartree atomic units ($m_e=e=\hbar=1$). For $E(t)$ in
Eq.~\ref{EqEt}, the two middle terms in Eq.~\ref{EqHamil} are given by
\cite{Hyper}
\begin{equation}
	V_P(\theta,t) = -\frac{1}{2} \Delta\alpha E_0^2 |f(t)|^2 \cos^2\theta
\end{equation}
\begin{eqnarray}\label{EqVhyper}
	V_H(\theta,t) &=& -\frac{3}{8}\beta_{xxz}E_0^3|f(t)|^3 \cos\theta \\ \nonumber
	              & & -\frac{1}{8}(\beta_{zzz}-3\beta_{xxz})E_0^3|f(t)|^3\cos^3\theta
\end{eqnarray}
where $\Delta\alpha = \alpha_\parallel - \alpha_\perp$ is the polarizability
anisotropy, and the $\beta_{ijk}$ are elements of the hyperpolarizability
tensor.  The molecular constants chosen to model CO are given in Table
\ref{TabConstants}.

Ionization is introduced using a complex absorbing potential
\begin{equation}
	V_I(\theta,t) = -(i/2) \Gamma_I(\theta,t)
\end{equation}
where $\Gamma_I(\theta,t)$ is the cycle-averaged ionization rate.  The
remaining time-dependence of $\Gamma_I(\theta,t)$ arises from the pulse
envelope $f(t)$.  The complex potential causes non-unitary quantum evolution
that removes amplitude as a function of angle, modeling the effects of
population loss due to ionization.
For simplicity, a separable form of $\Gamma_I(\theta,t)$ is assumed
\begin{equation}\label{EqSeparableIonization}
	V_I(\theta,t) = -(i/2)K(t) \Gamma_{\rm ref}(\theta),
\end{equation}
where $\Gamma_{\rm ref}(\theta)$ is the cycle-averaged angular ionization rate
calculated at a characteristic field strength $E_{\rm ref}$, and
\begin{equation}\label{EqTunExp}
	K(t) =  \exp\left(-\frac{2}{3}(2I_p)^{3/2}\left[|E_0f(t)|^{-1} - |E_{\rm ref}|^{-1}\right]\right) 
\end{equation}
is the tunneling exponent \cite{Landau,Keldysh} that provides the dominant
scaling of strong field ionization.  $I_p$ in Eq.~\ref{EqSeparableIonization}
is the ionization potential of the molecule.  Scaling in
Eq.~\ref{EqSeparableIonization} qualitatively captures the ionization yield in
small molecules over several orders of magnitude in laser
intensity \cite{SpannerChemPhys}.

\begin{table}[tbc]
    \caption{Molecular constants (a.u.) used to model CO.}
    \begin{ruledtabular}
    \begin{tabular}{cc|cc}
    Parameter & Value [Ref.] & Parameter & Value [Ref.] \\
    \hline
    $B$            & 8.7997$\times 10^{-6}$  \cite{NISTWebbook}     & 
    $\Delta\alpha$ & 3.6                     \cite{PetersonDunning} \\
    $\beta_{zzz}$  & 28.91                   \cite{PetersonDunning} &
    $\beta_{xxz}$  & 7.69                    \cite{PetersonDunning} \\
    $I_p$          & 0.516                   \cite{NISTWebbook}     &
    $c_0$          & 0.2214 $ \times 10^{-3}$  \\
    $c_1$          &-0.2141 $ \times 10^{-3}$  &
    $c_2$          & 0.0822 $ \times 10^{-3}$  
    \end{tabular}
    \end{ruledtabular}
    \label{TabConstants}
\end{table}

\begin{figure}[b]
	\centering
	\includegraphics[width=\columnwidth]{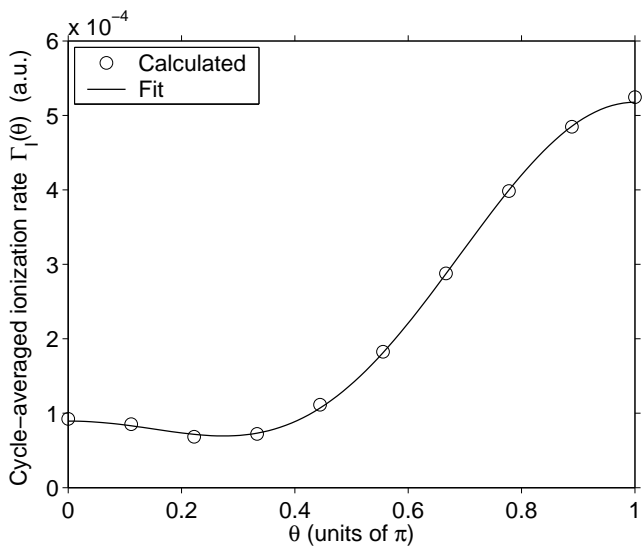}
	\caption{Cycle-averaged ionization rate $\Gamma_{\rm ref}(\theta)$
	Shown are the numerically calculated rates, and the fit from Eq.~(\ref{EqItheta}).}
	\label{FigVion}
\end{figure}

\begin{figure}[t]
	\centering
	\includegraphics[width=\columnwidth]{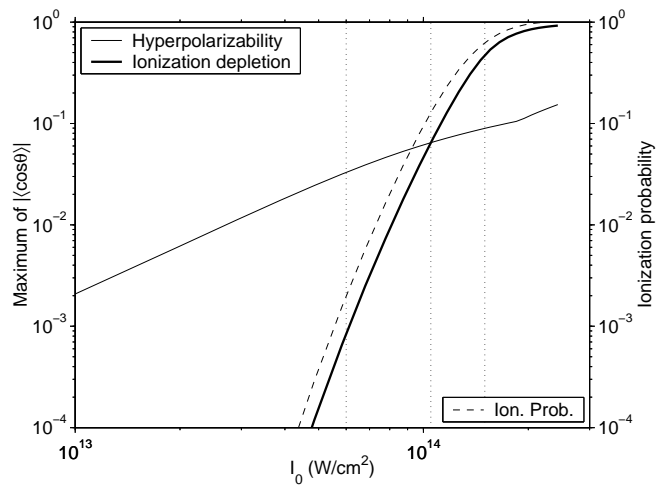}
	\caption{Left axis: the maximum field-free orientation ($|\langle\cos\theta\rangle|$) 
	after application of the two-color pump, as a function of pump intensity.  
	Each of the two colors in the pump pulse has intensity $I_0$.
	Thin line: the hyperpolarizability mechanism. Thick line:
	ionization depletion.  Right axis: total ionization probability (dashed line).
	The vertical dotted lines gives
	the intensities used in Figs.~\ref{FigCOhyper} and \ref{FigCOion}}
	\label{FigMaxCos}
\end{figure}

Calculation of $\Gamma_{\rm ref}(\theta)$ for ionization of X$^1\Sigma^+$~CO to
X$^2\Sigma^+$~CO$^{+}$ is carried out using the time-dependent mixed
orbital/grid method of Ref.~\cite{SpannerPatchkovskii}, in the single-channel
approximation.  The wavefunctions of the neutral and the ion are calculated
using GAMESS-US \cite{GAMESS} with the cc-pVTZ basis set \cite{Dunning} at the
complete active space level using 10/9 (neutral/cation) active electrons in 8
orbitals.  Uniform Cartesian grid extended to $\pm$13 Bohr, with spacing of 0.2
Bohr.  The time step was 0.002666 a.u.  The simulation was run for a full cycle
of the two-color field Eq.~(\ref{EqEt}), with $E_0 = E^*_0 = 0.0535$ ($I_0 =
10^{14}$ W/cm$^2$), $\omega = 0.057$ ($\lambda=$800/400 nm) followed by 2 fs at
zero field.  The cycle-averaged ionization rate was calculated by integrating
the flux absorbed \cite{Manolopoulos} at the edges of the grid, and dividing by
the cycle duration ($2\pi/\omega$).  Fig.~\ref{FigVion} plots the calculated
$\Gamma_{\rm ref}(\theta)$.  Also shown is the fit given by a truncated Fourier
series (see Table~\ref{TabConstants}):
\begin{equation}\label{EqItheta}
	\Gamma_{\rm ref}(\theta) = c_0 + c_1\cos\theta + c_2 \cos 2\theta
\end{equation}
All subsequent calculations use the fit of Eq.~\ref{EqItheta} in
Eq.~\ref{EqSeparableIonization}.

Although the Hamiltonian (\ref{EqHamil}) includes both the
hyperpolarizability and ionization terms, numerical results below
consider the two mechanisms individually, to elucidate
the characteristic features in the induced orientation.  
Simulations with both mechanisms active simultaneously
did not reveal any qualitatively new features, and will
not be discussed.

The initial populations of rotational states $|J,M\rangle$ are given by the
Boltzmann distribution at temperature $T=50$K:
\begin{equation}
	P(J,M) = \frac{\exp(-BJ(J+1)/kT)}{\sum_{J'}(2J'+1)\exp(-BJ'(J'+1)/kT)}
\end{equation}
The time evolution of each rotational state within the ensemble is expanded in
a spherical harmonics basis
\begin{equation}
	|\Psi(t)\rangle = \sum_J a_J(t) |J,M\rangle.
\end{equation}
The single sum over $J$ is appropriate since the Hamiltonian in
Eq.~(\ref{EqHamil}) conserves $M$.  The Schr\"odinger equation for the
coefficients $a_J(t)$ is
\begin{eqnarray}\label{EqSchro}
	i \frac{ \partial }{\partial t} a_J(t) &=& BJ(J+1) a_J(t) 
	\\ \nonumber
	&+& \sum_{J'} \langle J,M|V_P+V_H+V_I|J',M\rangle  a_{J'}(t).
\end{eqnarray}

The envelope function $f(t)$ is defined as 
\begin{equation}
	f(t) = \left \{
	\begin{array}{l l}
		0, & t<0 \\
		\sin\left({\pi t}/{2\tau_{on}}\right), & 0<t<2\tau_{on} \\
		0, & \:t>2\tau_{on} 
	\end{array} \right.
\end{equation}
corresponding to a $\sin^2$ pulse for the intensity $I=E^2$.  The parameter
$\tau_{on}$ ($\tau_{on}=30$~fs) is the full width at half-intensity.  While the
pulse is on ($t\le2\tau_{on}$), Eq.~(\ref{EqSchro}) is solved using the
Crank-Nicholson method \cite{NumericalRecipes}.  After the end of the pulse
($t>2\tau_{on}$), the analytical field-free propagation solution is used.
Each initial $|J,M\rangle$ state is propagated independently. The
observables are averaged over the thermal distribution $P(J,M)$.


\begin{figure}[t]
	\centering
	\includegraphics[width=\columnwidth]{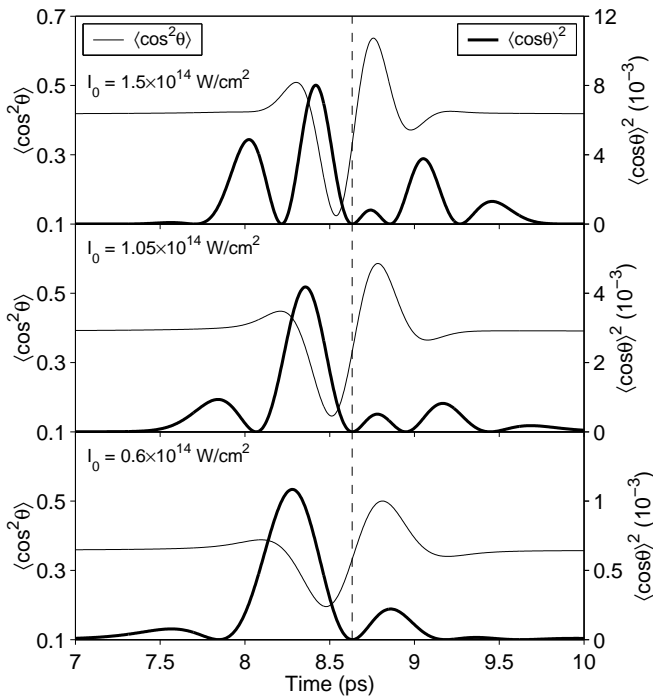}
	\caption{The hyperpolarizability orientation mechanism.
        Alignment (thin line, left axis) and orientation (thick line, right axis) wave-packet revivals
	for a selection of intensities. 
	The vertical dashed line denotes the revival time $t_{rev}$.}
	\label{FigCOhyper}
\end{figure}

\begin{figure}[t]
	\centering
	\includegraphics[width=\columnwidth]{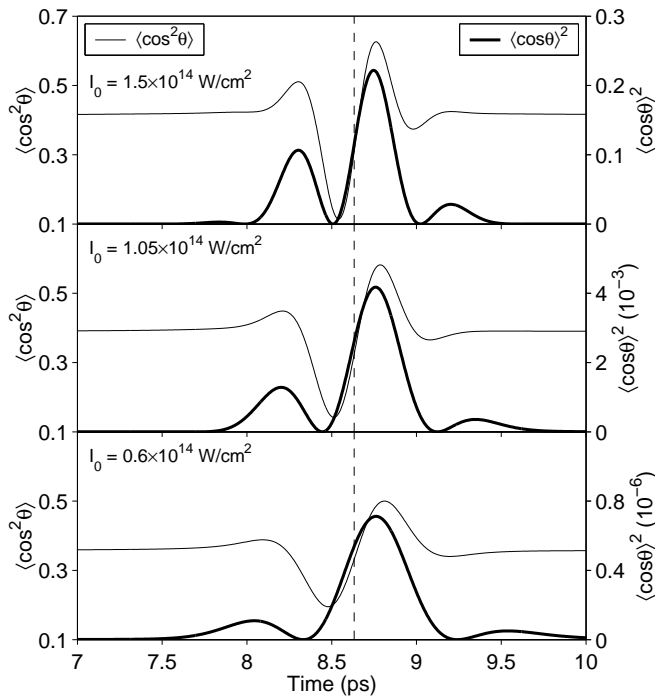}
	\caption{The ionization depletion mechanism. See Fig.~\ref{FigCOhyper}.}
	\label{FigCOion}
\end{figure}

The maximum field-free orientation following the pulse is shown in Figure
\ref{FigMaxCos}, as a function of the single-color intensity $I_0 = E_0^2$.
The orientation is characterized using the observable
$\langle\cos\theta\rangle$.  The total ionization probability is also shown
(dashed line, right axis) on the same figure.  The peak orientation scales very
differently with the intensity for the two mechanisms.  The hyperpolarizability
mechanism shows the expected $I_0^{3/2}$ scaling.  The depletion mechanism
shows the same scaling as the ionization probability, coming from the tunneling
exponent (Eq.~\ref{EqTunExp}).

At low intensities, where ionization is not possible hyperpolarizability
provides the dominant mechanism of orientation generation.  When the intensity
increases and substantial ionization occurs, ionization depletion dominates the
orientation.  For the present model, the cross-over occurs at $1.05\times 10^{14}$
W/cm$^2$.  This cross-over intensity is sensitive to the exact molecular parameters used,
especially to the ionization rate model.  Although qualitatively reasonable,
Eqs.~(\ref{EqSeparableIonization})-(\ref{EqItheta}) are not expected to be
quantitatively accurate, and consequently gives only a rough estimate of the
experimental cross-over intensity.  Due to the rapid increase in the peak
orientation once ionization is active, it is likely that ionization depletion
will be the active mechanism when the largest degrees of orientation are
reached, regardless of the intensity where the mechanisms switch.  Further,
since the peak orientation for the ionization mechanism parallels closely the
ionization probability, it follows that substantial orientation requires
substantial ionization. For example, reaching an orientation of
$\langle\cos\theta\rangle \gtrsim 0.1$ requires ionizing more than 10\% of the
sample in the CO case.

During short-pulse alignment and orientation, the molecular ensemble undergoes
wave-packet revival dynamics, where short periods of sharp angular localization
leading to strong alignment and orientation (the wave-packet revivals) are
separated by longer periods characterized by dispersed wavefunctions.  Figures
\ref{FigCOhyper} and \ref{FigCOion} show the alignment and orientation revivals
at the first full revival $t_{rev} = \pi/B \approx 8.64$ ps.  The alignment is
characterized by the commonly-used alignment parameter
$\langle\cos^2\theta\rangle$.  The orientation is tracked using
$\langle\cos\theta\rangle^2$, the square of the directional orientation
parameter $\langle\cos\theta\rangle$. Note that $\langle\cos\theta\rangle^2$
does not contain the orientation direction.  This is the appropriate observable
to characterize some experimental orientation probes, such as high-harmonic
generation \cite{EugeneCO}.  For the present study, this observable
conveniently draws out the peak orientation independent of direction.

At all intensities, the hyperpolarizability mechanism leads to peak
orientation at early times relative to $t_{rev}$
(Fig.~\ref{FigCOhyper}).  Since $t_{rev}$ is also typically near the center of
the alignment revival, the shift of the orientation wavepacket relative to
$t_{rev}$ can also be seen as a shift toward, and beyond, the minimum of the
alignment revival.  Ionization depletion, on the other hand, leads to
the peak orientation shifted to later times relative
to $t_{rev}$ and to the alignment revival (Fig.~\ref{FigCOion}).
Thus, the relative timing
between the peak of orientation and the alignment revival can be used to 
determine the active mechanism of orientation experimentally.

\begin{figure}[t]
	\centering
	\includegraphics[width=\columnwidth]{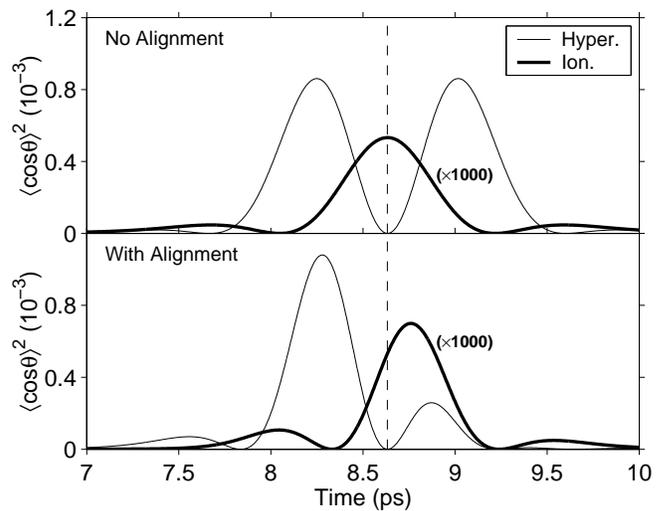}
	\caption{Orientation revivals within the impulsive approximation ($I_0=6.0\times 10^{13}$ W/cm$^2$).
	         The ionization depletion results are scaled by a factor of 1000.
	         Top panel: revival structure with the alignment interaction turned off ($\lambda_P=0$).
                 Bottom panel: with the alignment interaction included.
	The vertical dashed line denotes the revival time $t_{rev}$.}
	\label{FigCOexp}
\end{figure}

It is instructive to examine the origin of the relative timings within the two 
orientation mechanisms analytically, using the impulsive approximation.
If the pulse duration is much shorter
than typical rotation timescales ($\tau_{on} \ll 1/B$), the Hamiltonian 
of Eq.~\ref{EqHamil} becomes
\begin{equation}
	H = BJ(J+1) + \left[ V'_P(\theta) + V'_H(\theta) + V'_I(\theta) \right] \delta(t). \\
\end{equation}
where $ V'_i(\theta) = \int V_i(\theta,t) dt $.  Without sacrificing any of the
essential physics of the model, the potentials $V'_i$ can be approximated by
the leading non-constant terms in the $\cos\theta$ expansion
\begin{subequations}\label{EqJustCosCos2}
\begin{eqnarray}
	V'_P(\theta) &\approx& -\lambda_P\cos^2\theta \\
	V'_H(\theta) &\approx& -\lambda_H\cos\theta \\
	V'_I(\theta) &\approx& -i\lambda_I\cos\theta.
\end{eqnarray}
\end{subequations}
where the numerical parameters (at the single-color intensity $I_0 = 6.0\times 10^{13}$ W/cm$^2$)
are $\lambda_P = 3.823$, $\lambda_H = 0.2561$, and $\lambda_I = 1.095\times 10^{-3}$.
The $t=+0$ wavefunction then becomes \cite{Footnote}
\begin{equation}\label{EqDeltaPsi}
	|\Psi\rangle = \exp\left[ i\lambda_P\cos^2\theta + ( i\lambda_H - \lambda_I) \cos\theta \right]|\Psi_0\rangle.
\end{equation}
Impulsive orientation revivals for $I_0 = 6.0\times 10^{13}$ W/cm$^2$
and the initial ensemble temperature of 50K are shown in Fig.~\ref{FigCOexp}.
The impulsive approximation reproduces the temporal structure of the revivals
seen in the full simulations (cf. the bottom panel of Fig.~\ref{FigCOexp} to
the bottom panels of Figs.~\ref{FigCOhyper} and \ref{FigCOion}).  Thus, the
temporal signature of the orientation is unrelated to the finite pulse duration
effects.  Furthermore, the orientation revival timings are not affected by the
fine details of the angular dependence of the field-molecule interaction, and
are robust.

Consider now the orientation generated with alignment interaction turned off
(Fig.~\ref{FigCOexp}, top).  Both mechanisms generate revivals that are
completely symmetric around $t_{rev}$.  The hyperpolarizability revivals have a
zero at $t_{rev}$ and peak a short time after.  The ionization depletion
generates an orientation peak exactly at $t_{rev}$.  The alignment-free
behavior of the orientation is easy to understand.  The rotational wavepacket
is periodic, with period of $t_{rev}$.  Thus behavior at $t_{rev}$ is a
'mirror' of the initial motion.  Within the hyperpolarizability mechanism, the
molecules get an instantaneous kick from the $\delta(t)$ pulse, but the spatial
orientation distribution remains unchanged.  It takes a short time for the
imparted angular momentum to cause the molecules to rotate toward a point of
maximum orientation.  This is the exact behavior seen in Fig.~\ref{FigCOexp}.
Within the ionization depletion mechanism, the $\delta(t)$ pulse
instantaneously generates non-zero orientation by directly modulating the
angular distribution.  The instantaneously-generated orientation should then
disperse as the molecules start to rotate.  Again, this is clearly seen in
Fig.~\ref{FigCOexp}.

Including now the effects of alignment to the laser-induced orientation results
in the revival structures seen in the bottom panel of Fig.~\ref{FigCOexp}.  The
hyperpolarizability revival becomes asymmetric, but does not change position,
while the ionization depletion revival keeps its basic shape, but shifts to
later times.  The coupling of the alignment to the ionization depletion
revivals is qualitatively clear. The $\delta(t)$ pulse first creates
instantaneous orientation.  Then, as the $\cos^2\theta$ interaction generates a
torque that squeezes the angular distribution toward molecular alignment, the
instantaneously-generated orientation is also squeezed toward greater
orientation as the alignment maximizes (Figs.~\ref{FigCOhyper},
\ref{FigCOion}).  The coupling between the alignment dynamics and the
hyperpolarizability-generated orientation revival is not as obvious, but it is
still accounted for  by the coupling between the aligning and orienting forces.


Two mechanisms of laser-induced molecular orientation, due to
hyperpolarizability and ionization depletion, were studied using the CO
molecule as a model system.  At low intensities, the hyperpolarizability
mechanism dominates the generation of molecular orientation.  Once the ionizing
intensities are reached, the ionization depletion mechanism dominates.  The two
orientation mechanisms have clear parallels to the well-known sources of
vibrational wave packets: impulsive stimulated Raman scattering
(ISRS) \cite{SilvestriISRS}, and R-dependent ionization
(``lochfrass'' \cite{GollLochfrass}).  Indeed, the hyperpolarizability mechanism
acts by imparting angular momentum on the system, similar to ISRS.  The
ionization depletion on the other hand, directly ``burns-in'' the rotational
wave packet in the coordinate representation, analogously to lochfrass.  The
orientation revivals contain an unambiguous signature of the active orientation
mechanism.  the hyperpolarizability mechanism, the peak of the orientation
revival precedes the revival time $t_{rev}$, while for the ionization depletion
mechanism the peak of the orientation revival lags behind $t_{rev}$.  These
signatures arise due to interplay of the aligning and orienting interactions.



\begin{thebibliography}{}

\bibitem{StapelfeldtSeideman}
H. Stapelfeldt and T. Seideman, Rev. Mod. Phys. {\bf 75}, 543 (2003).

\bibitem{AttoReview}
F. Krausz and M. Ivanov, Rev. Mod. Phys. {\bf 81}, 163 (2009).

\bibitem{StolowAlign}
C.Z. Bisgaard, O.J. Clarkin, G. Wu, A.M.D. Lee, O. Ge{\ss}ner, C.C. Hayden, and A. Stolow, 
Science {\bf 323}, 1464 (2009).

\bibitem{Kling}
S. De, I. Znakovskaya, D. Ray, F. Anis, N.~G. Johnson, I.~A. Bocharova, M. Magrakvelidze, 
B.~D. Esry, C.~L. Cocke, I.~V. Litvinyuk, and M.~F. Kling,
Phys. Rev. Lett. {\bf 103}, 153002 (2009);

\bibitem{Hyper}
T. Kanai and H. Sakai, J. Chem. Phys. {\bf 115}, 5492 (2001).

\bibitem{Herschbach}
B. Friedrich and D. Herschbach, Phys. Rev. Lett. {\bf 74}, 4623 (1995).

\bibitem{NISTWebbook}
P.J. Linstrom and W.G. Mallard, Eds.,
NIST Chemistry WebBook, NIST Standard Reference Database Number 69,
National Institute of Standards and Technology,
Gaithersburg MD, 20899, http://webbook.nist.gov,
(retrieved September 25, 2011).

\bibitem{PetersonDunning}
K.~A. Peterson and T.~H. Dunning, Jr., J. Mol. Struct. (Theochem) {\bf 400}, 93 (1997).

\bibitem{Landau}
L.~D. Landau, E.~M. Lifshitz, 
{\it Quantum Mechanics: Non-Relativistic Theory}, 3$^{rd}$ edn. (Pergamon, Oxford 1977).

\bibitem{Keldysh}
L.~V. Keldysh, Zh. Eksp. Teor. Fiz. {\bf 47}, 1945 (1964)
[English transl.: Sov. Phys. JETP {\bf 20}, 1307 (1965)].

\bibitem{SpannerChemPhys}
M. Spanner and S. Patchkovskii, Chem. Phys. ({\it in press}, http://dx.doi.org/10.1016/j.chemphys.2011.12.016, 2012).

\bibitem{SpannerPatchkovskii}
M. Spanner and S. Patchkovskii, Phys. Rev. A {\bf 80}, 063411 (2009).

\bibitem{GAMESS}
M.W. Schmidt, K.K. Baldridge, J.A. Boatz, S.T. Elbert, M.S. Gordon, J.H.
Jensen, S. Koseki, N. Matsunaga, K.A. Nguyen, S. Su, T.L. Windus, M.
Dupuis, and J.A. Montgomery, J. Comput. Chem. {\bf 14}, 1347 (1993).

\bibitem{Dunning}
T.H. Dunning, Jr., Chem. Phys. {\bf 90}, 1007 (1989).

\bibitem{Manolopoulos}
D.E. Manolopoulos, J. Chem. Phys. {\bf 117}, 9552 (2002).

\bibitem{NumericalRecipes}
W.~H. Press, B.~P. Flannery, S.~A. Teukolsky, and W.~T. Vetterling,
{\it Numerical Recipes}, Cambridge University Press, Cambridge, 2$^{nd}$ edition, 1992.

\bibitem{EugeneCO}
E. Frumker, C.T. Hebeisen, N. Kajumba, J.B. Bertrand, H.J. W\"orner,  M. Spanner,
D.M. Villeneuve, A. Naumov, and P.~B. Corkum ({\it in submitted}, 2012).

\bibitem{Footnote}
This model accurately captures the temporal dynamics of the orientation, but no
longer includes the decay of the total wavefunction amplitude caused by
ionization.  The population decay can be recovered by restoring the
$\theta$-independent omponent of $V'_I(\theta)$ in Eq.~(\ref{EqDeltaPsi}).

\bibitem{SilvestriISRS}
S. de~Silvestri, J.G. Fujimoto, E.P. Ippen, E.B. Gamble Jr., L.R. Williams, and K.A. Nelson, 
Chem. Phys. Lett. {\bf 116}, 146 (1985).

\bibitem{GollLochfrass}
E. Goll, G. Wunner, and A. Saenz, Phys. Rev. Lett. {\bf 97}, 103003 (2006).


\end{thebibliography}
\end{document}